\documentclass[twocolumn,aps,prl,showpacs,amssymb,raggedbottom,nobalancelastpage,superscriptaddress]{revtex4}

\usepackage{amsmath}
\usepackage{amssymb}
\usepackage{amsfonts}
\usepackage{dsfont}
\usepackage{graphicx}
\usepackage{bm}
\usepackage{color}
\usepackage{appendix}
\usepackage{epsfig}

\begin{document}
\title{Topological States and Adiabatic Pumping in Quasicrystals}

\author{Yaacov E.~Kraus}
\affiliation{Department of Condensed Matter Physics, Weizmann Institute of Science, Rehovot, 76100, Israel.} %
\author{Yoav Lahini}
\affiliation{Department of Physics of Complex Systems, Weizmann Institute of Science, Rehovot, 76100, Israel.}
\author{Zohar Ringel}
\affiliation{Department of Condensed Matter Physics, Weizmann Institute of Science, Rehovot, 76100, Israel.} %
\author{Mor Verbin}
\affiliation{Department of Physics of Complex Systems, Weizmann Institute of Science, Rehovot, 76100, Israel.}
\author{Oded Zilberberg}
\affiliation{Department of Condensed Matter Physics, Weizmann Institute of Science, Rehovot, 76100, Israel.} %

\begin{abstract}
The unrelated discoveries of quasicrystals and topological insulators have in turn challenged prevailing paradigms in condensed-matter physics. We find a surprising connection between quasicrystals and topological phases of matter: (i) quasicrystals exhibit nontrivial topological properties and (ii) these properties are attributed to dimensions higher than that of the quasicrystal. Specifically, we show, both theoretically and experimentally, that one-dimensional quasicrystals are assigned two-dimensional Chern numbers and, respectively, exhibit topologically protected boundary states equivalent to the edge states of a two-dimensional quantum Hall system.We harness the topological nature of these states to adiabatically pump light across the quasicrystal. We generalize our results to higher-dimensional systems and other topological indices. Hence, quasicrystals offer a new platform for the study of topological phases while their topology may better explain their surface properties.
\end{abstract}


\maketitle

The discovery of topological insulators has sparked considerable interest in the study of topological phases of matter. Topological phases consist of various band insulators or superconductors that have gaps in their spectrum~\cite{RMP_TI}. The hallmark of these novel phases is the emergence of topologically protected boundary phenomena, e.g., quantum pumping \cite{Laughlin,QSHE}, surface states related to exotic models from particle physics \cite{Axions} and quasiparticles with non-Abelian statistics \cite{RMP_MF}. Yet, realizations of these phases of
matter are scarce~\cite{BiSe_Hasan,BiSe_Zhang,Heusler_Zhang,Heusler_Hasan,Demler_optics,Demler_PRA,Jonathan_PRB}.

Two systems belong to the same topological phase if they can be continuously deformed from one into the other without closing energy gaps. Consequently, at the interface between two topologically distinct systems, the energy gaps close by the appearance of localized boundary states. A classification of all the possible topological phases according to dimension and local symmetries was recently introduced~\cite{LudwigDim}. For example, in the absence of any symmetries,
all 1D systems belong to the topologically trivial phase, while in 2D there are the topological phases of the integer quantum Hall effect (IQHE)~\cite{Avron1}.

The order of quasicrystals (QCs) -- nonperiodic structures with long-range order -- can be seen as originating from periodic structures of a dimension higher than the physical one. For example, the 1D Fibonacci QC can be described as a projection of a 2D lattice on a line~\cite{QC_Janot}. Remarkably, observed phenomena such as unconventional Bragg diffraction and the existence of phasons can be attributed to this higher dimension~\cite{Phasons,QC_Janot,Freedman}. Remnants of the higher dimensionality appear as additional degrees of freedom (d.o.f.) in the form of shifts of the origin of the quasiperiodic order. These d.o.f. discern between QCs with the same quasiperiodic order, as they result in different patterns. However, they have no apparent influence on bulk properties and were therefore usually ignored.

In this Letter, we show that, due to the additional d.o.f.,
QCs exhibit nontrivial topological properties that are attributed
to systems of a higher dimension. The topological
properties of the QC manifest in two ways: (i) the existence
of quantum phase transitions when continuously deforming
between two topologically distinct QCs and (ii) the appearance
of robust boundary states which traverse the bulk gaps
as a function of the aforementioned shifts. Specifically, we
demonstrate, both theoretically and experimentally, that 1D
QCs exhibit topological properties that were, thus far,
thought to be limited to 2D systems. Using photonic QCs,
we observe localized boundary states, which manifest these
topological properties. The topological nature of these
boundary states is used to realize an adiabatic pumping of
photons across the sample. Generalizations to various types
of quasicrystals in 1D and higher dimensions are also discussed,
suggesting the existence of topological effects on
surfaces of 3D quasicrystals.

\begin{figure}
\includegraphics[clip,scale=0.55,width=\columnwidth]{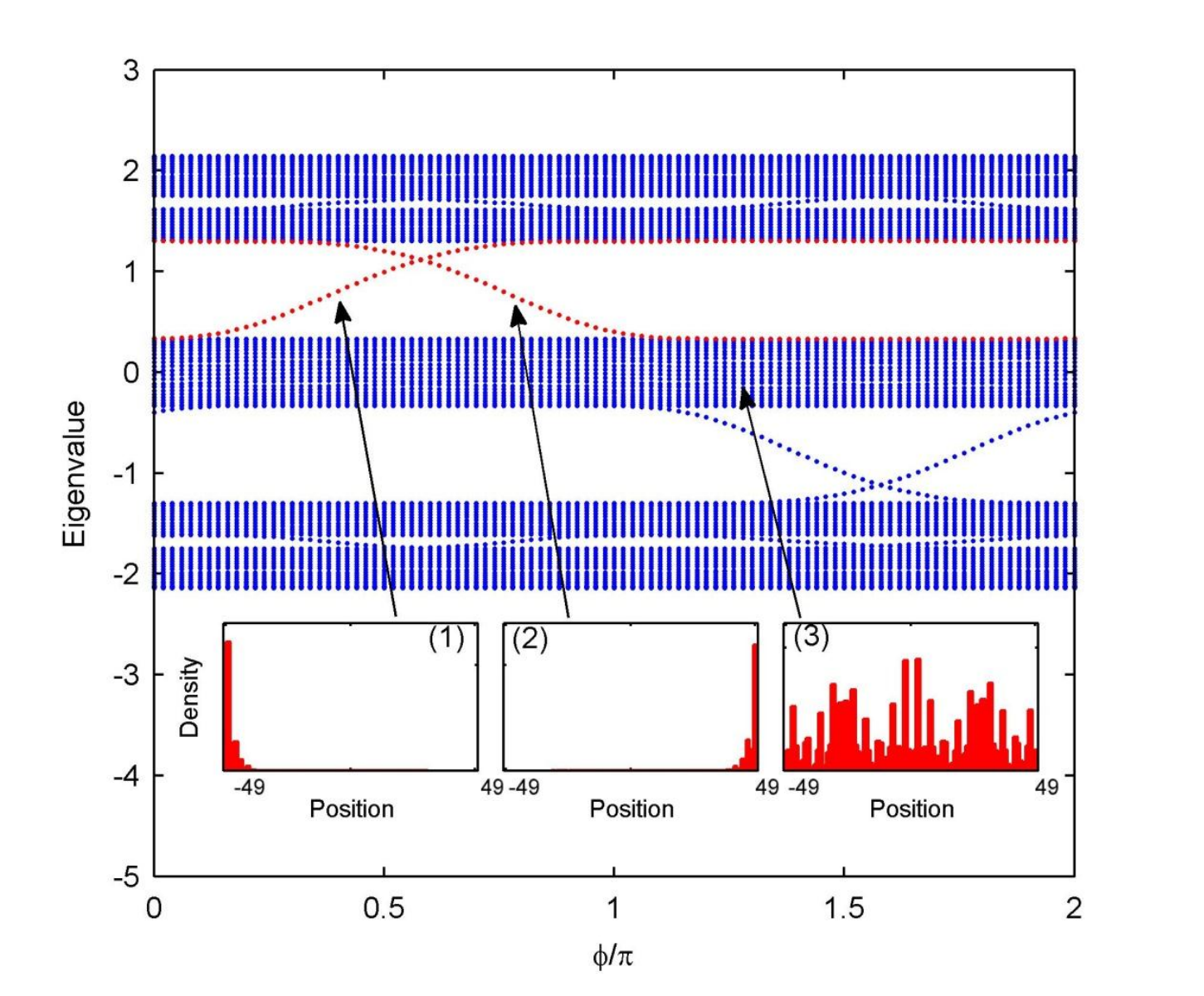}%
\caption {\label{Fig1}  The numerically calculated spectrum of Eq.~\eqref{eq:H_phi}
as a function of the phase $\phi$ for $t=1$, $\lambda=0.5$,
$b=\left(1+\sqrt{5}\right)/2$ (the Golden mean), and $n=-49...49$. The bulk of the spectrum remains fixed, whereas few modes, localized at the boundaries, sweep across the gaps. The insets depict
the spatial density of typical eigenstates as a function of position
along the 1D lattice: (1) a left boundary state, (2) a right
boundary state, and (3) an extended state within the band.}
\end{figure}

Let us begin with a specific QC, the 1D Aubry-Andr\'{e} (AA)
model~\cite{AA} (also known as the Harper model). This is
a 1D tight-binding model in which the on-site potential is
modulated in space. It is described by the Hamiltonian
\begin{equation}
H(\phi)\psi_{n}=t(\psi_{n+1}+\psi_{n-1})+\lambda\cos\left(2\pi
bn+\phi\right)\psi_{n}\,.\label{eq:H_phi}
\end{equation}
Here, $\psi_{n}$ is the wavefunction at site $n$, $t$ is the hopping amplitude,
$\lambda$ is the modulation amplitude of the on-site potential, and $b$ controls the
periodicity of the modulation. Whenever $b$ is irrational the modulation is
incommensurate with the lattice and the on-site term is quasi-periodic. Note that
in this model the modulation phase $\phi$ embeds the d.o.f.~mentioned above.

Figure \ref{Fig1} depicts a numerically calculated spectrum of the AA model as a
function of $\phi$. Because of the incommensurate potential, the spectrum is broken into a fractal set of bands and gaps~\cite{Hofstadter}. Note that as a function
of $\phi$ the bands are almost unchanged, but the gaps are crossed by a
few modes. These states that reside within the gaps are
boundary states, localized either on the left or on the right
boundary of the system, as seen in insets (1) and (2),
respectively. The states within the bands are typically
extended, as depicted in inset (3). As we later show, these
boundary states are the physical manifestation of the fact
that the AA model belongs to a nontrivial topological
phase.

We implemented the AA model in an optical setup using
a quasiperiodic lattice of coupled single-mode waveguides.
Because of a nonvanishing overlap between the evanescent
modes, light that propagates along a waveguide can hop to
its neighboring waveguides. In addition, along each waveguide
a phase is accumulated in a rate determined by its
refraction index. Hence, the propagation of light along
the lattice is described by an equation which is identical
to a tight-binding model, where the propagation axis, \emph{z}, takes over the role of time, $i\partial_z\psi_n = H\psi_n$. Modulating the
refraction index of the waveguides and the spacing between
them controls the on-site and hopping terms of the
Hamiltonian, respectively~\cite{WGA,Szameit,Lahini}. In particular, it enables
a direct realization of the AA Hamiltonian of Eq.~\eqref{eq:H_phi}.

\begin{figure}[htb]
\includegraphics[clip,scale=0.49]{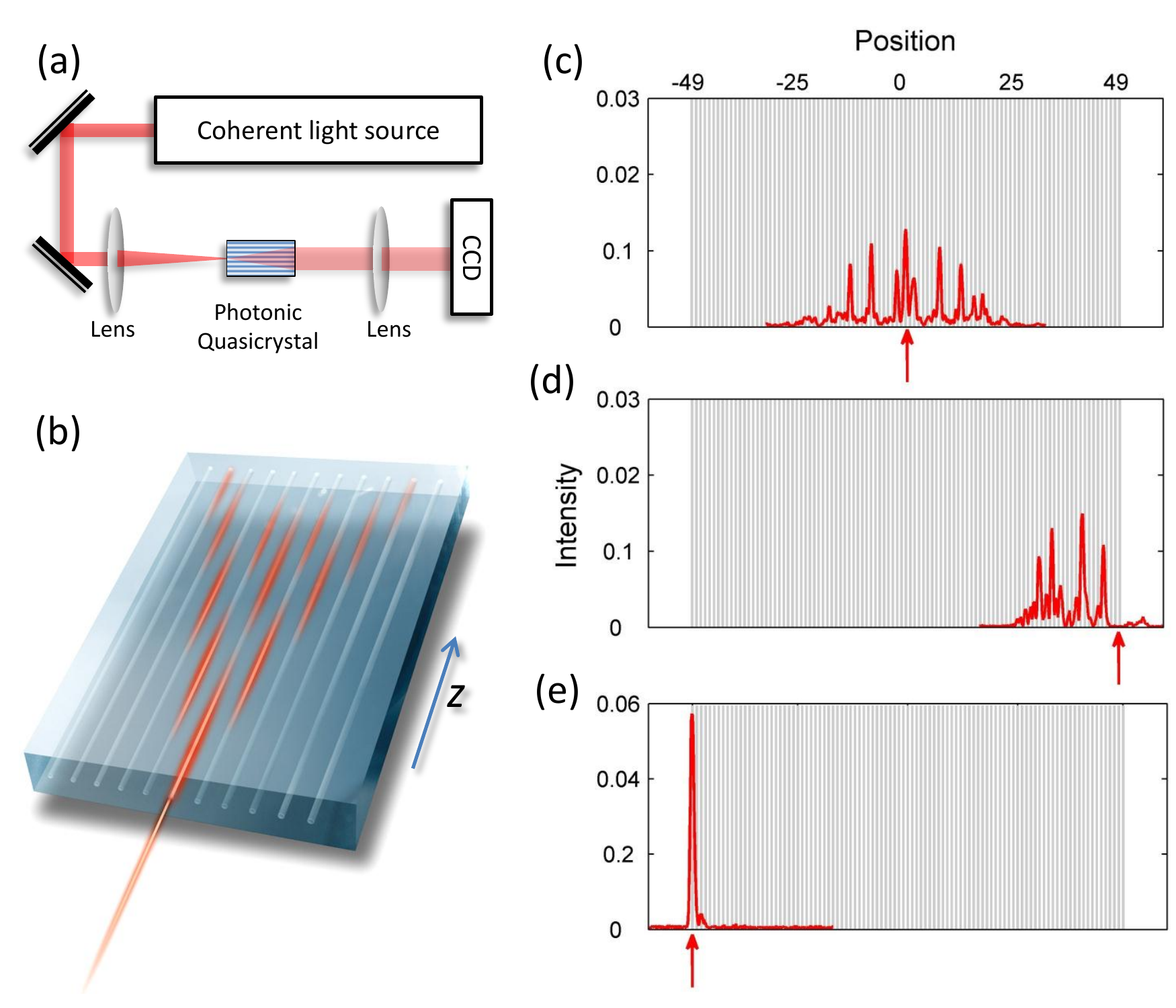}%
\caption {\label{Fig2} Observation of topological boundary states in an
Aubry-Andr\'{e} photonic quasicrystal. (a) A sketch
of the experimental setup. (b) An illustration of the conducted
experiment. Light is injected into one of the waveguides and
tunnels to neighboring waveguides as it propagates. (c)–(e) Experimental observation of the left boundary state for
$\phi=\pi/2$. Light was initially injected into a single waveguide (red
arrows). The measured outgoing intensity is plotted versus the
injection position along the lattice. (c),(d) An excitation at the
middle of the lattice (site $0$) and at the rightmost site (site $49$)
results in a significant spread. (e) For an excitation at the leftmost
site (site $-49$), the light remains tightly localized at the boundary,
marking the existence of a boundary state.}
\end{figure}

We produced an AA lattice with the parameters of Fig.~\ref{Fig1} and $\phi=\pi/2$ on
a semiconductor (ALGaAs) substrate using standard photolithography
methods~\cite{supmat}. We set the effective refraction index of each waveguide -- by
controling its width -- to fit the prescribed quasi-periodic pattern. We injected light into a single waveguide and measured the outgoing intensity distribution, as illustrated in Figs.~\ref{Fig2}(a) and \ref{Fig2}(b). The experimental observations are depicted in Figs.~\ref{Fig2}(c-e). Light injected into a lattice site
in the middle of the lattice showed a significant expansion,
due to the overlap of the injected wave function with the
extended bulk eigenstates. Similarly, light injected into the
rightmost lattice site showed considerable expansion.
However, when the light was injected into the leftmost
lattice site [see Fig.~\ref{Fig2}(e)], the intensity distribution remained
tightly localized at the boundary, with the
maximum intensity found at the leftmost waveguide, itself.
This is a clear signature of the existence of a localized
boundary state.

\begin{figure*}[htb]
\includegraphics[clip,scale=0.7]{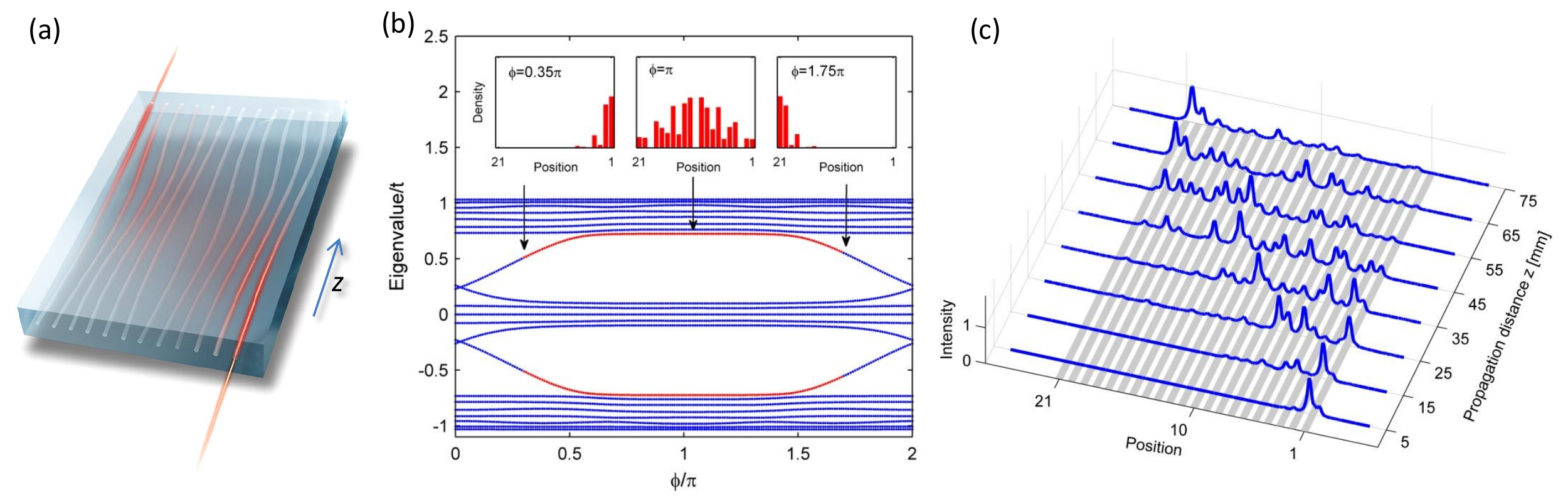}
\caption {\label{Fig3} Experimental observation of adiabatic pumping via topologically protected boundary states in a photonic
quasicrystal. (a) An illustration of the adiabatically modulated photonic quasicrystal, constructed by slowly varying the spacing
between the waveguides along the propagation axis $z$.
Consequently, the injected light experiences an adiabatically modulated
Hamiltonian, $H_\text{off}(\phi(z))$, as it propagates and is pumped across the sample. (b) The spectrum of the model as a function of the phase
$\phi$ for $t=40/75$, $\lambda=0.6$, $b=\left(1+\sqrt{5}\right)/2$, and $n=1...21$.
In the experiment, $\phi$ was scanned between $0.35\pi$ and $1.75\pi$, marked by
arrows (and red dots). The insets depict the spatial density of a boundary eigenstate as a function of the position at three different stages
of the evolution: At $\phi=0.35\pi$, the eigenstate is localized on the right boundary. At $\phi=\pi$, it is delocalized across the system, while at $\phi=1.75\pi$ the state is
again localized, but on the left boundary. (c) Experimental results: Light was injected into the rightmost waveguide (site $1$) at $z=0$ ($\phi=0.35\pi)$.
The measured intensity distributions as a function of the position are presented at different stages of the adiabatic evolution, i.e., different propagation distances. It is evident that along the adiabatic evolution the light crossed
the lattice from right to left.}
\end{figure*}

A consequence of the topological nature of this model is
that all boundary states which reside within the same gap
belong to the same mode. This can be seen by following the
eigenenergy of some boundary state as a function of $\phi$. For
example, take the right boundary state denoted in inset (2)
of Fig.~\ref{Fig1}. Its mode is marked by red circles. This state
remains localized on the right boundary as long as its
energy remains within the gap. When the energy reaches
the band, the state becomes extended. Notably, once the mode returns into the gap, it appears localized at the opposite boundary.

This property was used to realize adiabatic pumping of
photons from one side of the lattice to the other. A convenient
platform for this feat is the ``off-diagonal''
version of the AA model, which is described by the Hamiltonian
\begin{align}
\label{off_AA}
H_{\text{off}}(\phi)\psi_{n} &= t\left[1+\lambda\cos\left(2\pi bn+\phi\right)\right] \psi_{n+1} \\
& + t\left[1+\lambda\cos\left(2\pi b(n-1)+\phi\right)\right] \psi_{n-1}\,. \nonumber
\end{align}
While this model embeds its quasiperiodicity in the hopping
term, it has topological characteristics similar to its previously
discussed ``diagonal'' version [cf.~Eq.~\eqref{eq:H_phi}]. The pumping takes place when $\phi$ is adiabatically swept along the propagation axis $z$.

In our implementation, we used waveguides written in
bulk glass using femtosecond laser microfabrication technology~\cite{Szameit}. The spacing between the waveguides was
slowly modified along the propagation axis, thus realizing
a sweep of $\phi$ in
Eq.~\eqref{off_AA} [see Fig.~\ref{Fig3}(a)]. The length of the
sample was $75$\textit{mm}, which is in our case $20$ tunneling
lengths, where the tunneling length is the characteristic
scale for hopping, namely, $2/t$~\cite{supmat}. Figure \ref{Fig3}(b)
depicts the spectrum of the system as a function of $\phi$. In order to
observe different stages of the pumping process, we fabricated
a set of $50$ samples for which the light was allowed
to propagate shorter distances within the modulation.
Correspondingly, in the $i^\text{th}$ sample, $\phi$ is modulated from $0.35\pi$ to $(0.35+1.4\cdot i/50)\pi$. For each sample, light was
injected to the rightmost site and the output intensity distribution was measured. The collected results are summarized
in Fig.~\ref{Fig3}(c). The obtained intensity distributions
are stacked incrementally according to their propagation
distance, i.e., their final $\phi$. Thus, we reconstruct the light's
trajectory along the full adiabatic process. It is evident that
the injected light was pumped adiabatically across the QC
from one boundary to the other~\cite{pumping}.

\vspace{5mm}

We now turn to establish theoretically the topological
properties of QCs. We start by showing that the observed
boundary states are of topological origin by mapping the
AA model to the lattice version of the 2D IQHE~\cite{Hofstadter}. In the
latter, electrons hop on a 2D rectangular lattice with
nearest-neighbor hopping amplitudes $t$ and $t'$ in the presence
of a perpendicular magnetic field, with $b$ flux quanta
threading each rectangle. Assuming one coordinate to be
periodic and using the Landau gauge for the magnetic field,
the system can be described by the Hamiltonian
$\mathcal{H}\psi_{n,k}=\sum_{k}\left[t(\psi_{n+1,k}+\psi_{n-1,k})+2t'\cos(2\pi
bn+ka)\psi_{n,k}\right]$, where $k$ is
the momentum along the periodic coordinate with lattice
spacing $a$ and $n$ is the location in real space along the
second coordinate. The energy spectrum of $\mathcal{H}$ is gapped,
and each gap is associated with a quantized Hall conductance $\sigma_{H}=\nu e^{2}/h$, with $\nu$ an
integer~\cite{TKNN} known as the Chern number~\cite{Kohmoto_AnnPhys}. The inclusion
of disorder and distortions in the Hamiltonian does not alter $\sigma_{H}$, as long
as the corresponding gap is maintained open~\cite{TKNN,Avron0,Kohmoto_AnnPhys}. Because of the fact that the energy gap must be closed in order for $\sigma_{H}$ to change its value, it can be used to classify different
phases of the IQHE. Phases with different $\sigma_{H}$ are said to be topologically distinct.

The physical manifestation of a nontrivial topological phase
(i.e., $\sigma_{H}\neq0$) is the emergence of robust chiral
states along the edges of the sample. This general phenomenon
is shared by many topological phases, not only
the IQHE~\cite{Hatsugai,KaneMele_Z2,ZoharKobi}. In the IQHE, on each edge, exactly
$|\nu|$ edge states appear, and the energy of each edge state
traverses the gap as
$k$ varies from $-\pi/a$ to $\pi/a$. The signs
of the group velocity of these edge states are opposite on
opposite edges.

Turning back to the 1D AA model [cf.~Eq.~\eqref{eq:H_phi}], we shall now
observe that it inherits its robust boundary states from the 2D IQHE. For
$\lambda=2t'$, the spectrum of $H(\phi)$ can be viewed as the $k^{\textrm{th}}$
component of $\mathcal{H}$ at $k=\phi/a$. Therefore, by scanning $\phi$ from $-\pi$
to $\pi$, the spectrum of $H(\phi)$ reconstructs that of $\mathcal{H}$.
Consequently, the chiral
edge states that traverse the gaps as a function of $k$
appear now as 1D boundary states that traverse the gaps with $\phi$. Since these boundary states are of topological origin, the
only way to eliminate them is to close the energy gap
which they traverse. In particular, disorder which does
not close an energy gap does not eliminate the corresponding
boundary states.

Note that the topology guarantees the existence of
boundary states only for intervals of $\phi$. Hence, it does
not guarantee that for any QC pattern they indeed appear.
This can be seen, for example, in Fig.~\ref{Fig3}(b), where for
$\phi=0$ one finds states localized on both boundaries and for $\phi=\pi$ there are
none. Additionally, changing the number of lattice
sites or translating them (e.g., taking $n=5...25$ instead of $n=1...21$) may alter dramatically the intervals
of $\phi$ for which boundary states appear.

So far, in order to witness the topological nature of the
AA model, we had to scan $\phi$. Ostensibly, one can associate
Chern numbers only to the union of all the $H(\phi)$ Hamiltonians. Indeed, per $H(\phi)$, a Chern density is assigned,
while the Chern number involves integration of the
Chern density over all $\phi$. However, we show that for QCs
the Chern density is independent of $\phi$. Thus, the Chern
number can be evaluated from the Chern density of any $H(\phi)$. Since the same quantized Chern number is associated
with any $H(\phi)$, it topologically classifies it. This is
somewhat analogous to the role of the Aharonov-Bohm
flux in the IQHE~\cite{supmat,TKNN,ZoharKobi}. The association of a Chern
number with each QC is a key result of this work.

Proof of the above statement appears in the
Supplemental Material~\cite{supmat}.
Here, we show that the bulk spectrum is also independent of $\phi$. This simpler proof contains the essential ingredients of the former. Since $H(\phi)$ has a band structure, the spectrum is insensitive to
lattice translations in the thermodynamic limit. From
Eq.~\eqref{eq:H_phi}, it is evident that translating the lattice by
$m$ sites is equivalent to shifting $\phi$ by $2\pi\cdot(bm$ mod$1)$. Now the
irrationality of $b$ comes into play. For a rational $b=p/q$, $(bm$ mod$1)$ has only
$q$ different values for all possible translations. Thus, the band structure is
guaranteed to be invariant only for these $q$ corresponding shifts of $\phi$. On the
other hand, for irrational $b$, $(bm$ mod$1)$ samples the entire $[0,1]$ interval
and the bands are invariant for any shift of $\phi$.

Thus, we arrive at the following conclusion: while in
order to witness boundary effects the scanning over $\phi$ is
required, the topological indices can be associated with any
instance of a quasiperiodic pattern, i.e., any given $\phi$.
These indices are, of course, the same for a given quasiperiodicity
for all $\phi$'s. Thus, the AA model is topologically classified.
Consequently, two QCs with two different $b$'s cannot be
smoothly deformed from one to the other without closing
the bulk gaps, since in the IQHE different $b$'s result in different Chern
numbers~\cite{supmat}.

Until now, we focused on a specific model which we
were able to map to the IQHE. However, our results could
be easily generalized to any QC, such as the off-diagonal
AA model and the Fibonacci QC. Moreover, our arguments
apply to any dimension and any topological index without
the need for establishing such a mapping~\cite{supmat}. Consider a $D$-dimensional QC with a tight-binding
Hamiltonian with $d$ quasiperiodic terms, either hopping or on-site. These terms
result in $d$ d.o.f.~similar to the above $\phi$. In the context
of topological properties, these d.o.f.~could be treated as
extra dimensions, yielding an overall effective dimension
of $D+d$. Therefore, the Hamiltonian may belong to a
nontrivial $D+d$-dimensional topological class.

We showed that quasicrystals exhibit new types of topological
phases that were previously attributed only to
systems of higher dimension. The study of these novel
topological phases in 2D and 3D may lead to the discovery
of new surface phenomena in atomic and photonic quasicrystals;
e.g., 3D quasicrystalline materials may exhibit
topological properties that would have appeared only in
6D periodic systems. Furthermore, our approach provides
new tools for engineering photonic quasicrystals and especially
for controlling their surface properties.

We thank Y.~Silberberg, S.~Huber, Y.~Gefen, and E.~Altman for useful discussions and N.~Gontmakher for the illustration of the device. We especially thank Y.~Silberberg for allowing us to conduct the experiments in his labs. We
thank the U.S.–Israel Binational Science Foundation, the
Minerva Foundation, Crown Photonics Center, ISF Grant
No. 700822030182, and the IMOS Israel-Korea Grant for
financial support. All authors contributed equally to this
work.

\newpage\begin{center}
\textbf{\large SUPPLEMENTAL MATERIAL}
\end{center}

\section{Experimental setups}
\label{Sec:Experiment}

The main text discusses a new theoretical finding: quasiperiodic systems are
topologically nontrivial and, correspondingly, have eigenstates within the bulk gap
that reside at the boundaries. It also describes two experiments that confirm the
existence of these boundary states and their topological nature. In order to perform
these experiments we have fabricated 1D photonic quasicrystals that constitute exact
realizations of the tight-binding Hamiltonians appearing in the main text.

\begin{figure}[htb]
\includegraphics[width=0.8\columnwidth]{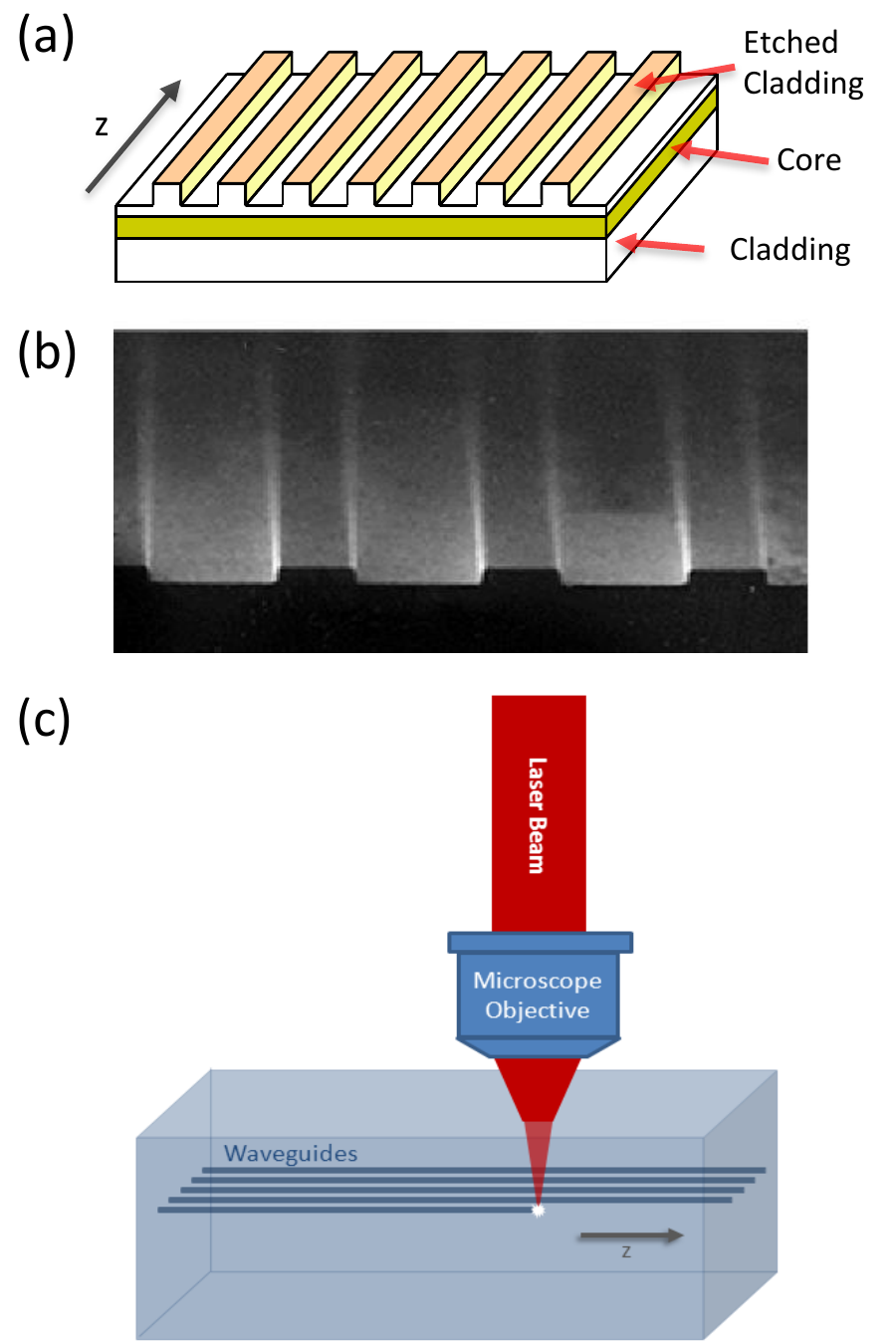}%
\caption {\label{Fig1appendix}  The two types of waveguide lattices used in the two
experiments. (a) An AlGaAs waveguide lattice, composed of three layers:
cladding-core-cladding. The etching of the upper layer controls locally the
effective refraction index of the middle layer. Quasiperiodic etching of the
cladding realizes the AA model in the core layer. (b) A SEM image of the fabricated
AlGaAs waveguide lattice. (c) Waveguides inscribed into bulk glass using a
femtosecond-laser writing technique. }
\end{figure}

The first set of experiments showed the existence of boundary states by studying the
expansion of injected light into different entry points. It was conducted using an
array of evanescently coupled waveguides fabricated on a semiconducting AlGaAs
substrate. The array is composed of a core layer sandwiched between two cladding
layers, where the upper cladding layer is etched quasiperiodically, see
Figs.~\ref{Fig1appendix}(a) and \ref{Fig1appendix}(b). The etching makes the core beneath it have a
lower effective refraction index, resulting in a array of coupled 1D waveguides.

In order to realize the AA model [cf.~Eq.~(1) of the main text], we modulated
quasiperiodically the effective refraction index of the waveguides, by changing
their width while keeping the inter-waveguide separation constant. The mean width of
the waveguides was 3$\mu m$, and the separation was 9$\mu m$. The three layers were
1.5, 1.5 and 4$\mu m$ thick, and the etching depth was of 1.1$\mu m$. The core layer
was of 18\% aluminum mole-fraction core, while the cladding layers was of 24\%. The
sample length was 20$mm$, which is equivalent to 30 tunneling lengths, where the
tunneling length is the characteristic scale for hopping, namely $2/t$. The light
source was a continuous-wave laser diode with a wavelength of 1550$mm$. The laser
beam was focused into a single waveguide at the input facet of the array using a
$\times$40 microscope objective. The light at the output facet was imaged onto an
infrared camera (Hamamatsu IR C5840).

The second experiment utilized the topological nature of the boundary states for
performing adiabatic pumping of light. For this experiment we used several arrays of
waveguides written in bulk glass using femtosecond-laser microfabrication
technique~\cite{Szameitappendix}, see Fig.~\ref{Fig1appendix}(c). The waveguides were all identical
in both refraction index and width (2$\mu m$), while the inter-waveguide separation
was modulated in order to realize the off-diagonal AA model [cf.~Eq.~(2) of the main
text]. The sample length was 75$mm$, which is equivalent to 20 tunneling lengths.
The number of waveguides in each array was 21. The light source was a
continuous-wave laser diode with a wave-length of 808$nm$.

\begin{figure}[htb]
\includegraphics[width=\columnwidth]{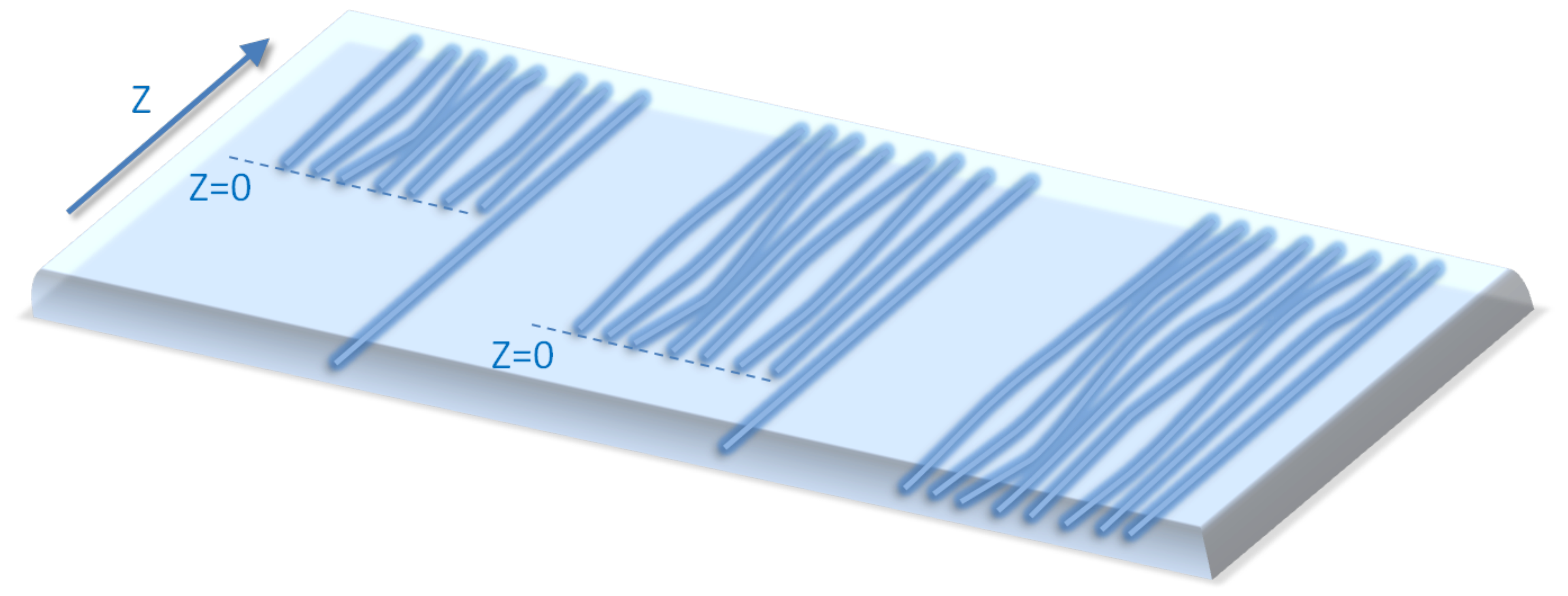}%
\caption {\label{pumping}  An illustration of arrays of waveguides written in the
bulk glass used to measure the wavefunction evolution ``inside'' the lattice during
the adiabatic pumping. The rightmost array realizes the full pumping, while the
others realize only parts of it. }
\end{figure}

The inter-waveguide separation was modulated in the slowest possible rate, for which
the light successfully pumps as it propagates through the whole sample length. In
such an array, when light was injected to the rightmost waveguide, the outgoing
light distribution was measured to be localized at the leftmost waveguide. Notably,
during the adiabatic evolution, the light is expected to first expand across the
array and only later to be collected at the opposite side. In our setup, however, we
were able to measure the light intensity distribution at the output only. Therefore
the observation of the initial expansion seemed impossible.

In order to, nevertheless, measure the light distribution during the pumping
process, we produced a set of arrays with the same modulation rate. Each array
realized only a part of the full adiabatic evolution. This is illustrated in
Fig.~\ref{pumping}, where only the rightmost array realizes the full pumping. The
adiabaticity ensures that for a given propagation length, the outgoing light of a
partial-evolution is distributed in the same way that it would have been distributed
in the midst of the full evolution. Therefore, by measuring the output of lattices
with partial-evolution we, in fact, measure the wavefunction ``inside'' the lattice
during the full adiabatic pumping. Notably, the partial evolutions are shorter than
the full one, but the sample length is fixed. Therefore, we used a single waveguide
lead that delivered the light to the beginning of the waveguide array, as depicted
in Fig.~\ref{pumping}.

\section{The Chern number of a quasicrystal}
\label{Sec:Chern}

In this section we show that Chern numbers can be defined for any given 1D
quasicrystal (QC). This is accomplished by proving that the Chern density associated
with a gap of the QC is independent of the phase of the quasiperiodicity in the
thermodynamic limit. The proof is presented for the Aubry-Andr\'{e} (AA) model
[cf.~equation (1)], but can easily be generalized to other QCs and topological
indices.

In order to evaluate the Chern number we consider a modified AA model
with a similar spectrum. This model describes a periodic 1D lattice
(a ring) of length $L$ with an on-site potential parameterized by
an irrational number $b$, and the Hamiltonian
\begin{align}
H_b(\phi,\theta)\psi_n = & t e^{i\theta/L}\psi_{n+1}+te^{-i\theta/L}\psi_{n-1} \nonumber \\
 & + \lambda\cos\left(2\pi\bar{b}_L n+\phi\right)\psi_n \, ,
\end{align}
where $\bar{b}_{L}=\lfloor b \cdot L\rfloor / L$ is a rational
approximation to $b$ that makes the modulation periodic. It is clear that for
$\theta=0$ and $L\rightarrow\infty$ this Hamiltonian coincides with the one of the
AA model. For finite but large $L$, the spectrum of $H_{b}$ has the same gaps as
those of the AA model, up to small corrections.

The phase factor, $\theta$, in the hopping terms introduces a phase twist along the
ring. Equivalently, it represents a magnetic field of $\theta/2\pi$ flux quantum
that threads through the ring. It plays a role similar to the role of the momentum
in a translation invariant crystal, and is required in order for the Chern number to
be defined~\cite{Niu_Thouless}.

For an integer $\theta/2\pi$ this phase factor can be gauged out. Hence, the
spectrum and the topological properties are periodic in $\theta$ modulus $2\pi$.
Moreover, since $\theta$ appears in the Hamiltonian only via $\theta/L$-terms,
perturbation theory shows that its influence on the spectrum is suppressed by a
factor of $1/L$. Therefore, for large enough $L$ the gaps of $H_{b}$ are well
defined for every $\theta$. In the thermodynamic limit the effect of $\theta$
completely vanishes.

We can therefore summarize that a gap of $H_{b}$ that remains open while $L$ is
increased corresponds one-to-one to a gap of the AA model. Consequently, such a gap
has, in the thermodynamic limit, the same physical properties in both models, in
particular the same Chern number.

Consider the shift $\phi \rightarrow \phi + \epsilon$, where $\epsilon=2\pi l/L$ and
$l=0,1,...,L-1$. There is always some translation of the lattice sites $n
\rightarrow n + n_\epsilon$, where $n_{\epsilon}\in0,1,...,L-1$, such that
$\cos\left(2\pi\bar{b}_{L}n+\phi+\epsilon\right)=\cos\left(2\pi\bar{b}_{L}(n+n_{\epsilon})+\phi\right)$.
Note that $n_\epsilon$ is independent of $\phi$. For a prime $L$ it is also
guaranteed that there is an $n_{\epsilon}$ corresponding to the minimal
$\epsilon=2\pi/L$. Therefore, if we denote by $\hat{T}_{\epsilon}$ a translation
operator by $n_{\epsilon}$ sites, then
$H_{b}(\phi+\epsilon)=T_{\epsilon}H_{b}(\phi)T_{\epsilon}^{-1}$. The equivalence
between the phase shifts and spatial translations implies that the shifts have no
physical consequences, such as closing of the energy gaps. In the thermodynamic
limit $\epsilon$ becomes continuous, which means that the spectrum is independent of
$\phi$, as mentioned in the main text. For finite but large enough $L$ the spectrum
weakly depends on $\phi$ and the gaps do not close as a function of $\phi$.

A convenient way to evaluate the Chern number associated with a given gap is
to consider the projector on the states below this gap
\begin{align}
P(\phi,\theta) & =\sum_{E_{n}<E_\text{gap}}|n\rangle\langle n|\,,
\end{align}
where $|n\rangle$ is an eigenstate of $H_b$ with energy $E_n$, and $E_\text{gap}$ is
the energy at the center of the gap. For large enough $L$, such that the gaps are
open for any $\phi$ and $\theta$, $P(\phi,\theta)$ is a well defined quantity. Since
the projector differs from the Hamiltonian only in its eigenvalues, we similarly have
$P(\phi+\epsilon)=T_{\epsilon}P(\phi)T_{\epsilon}^{-1}$. Moreover, by definition
\begin{align}
\partial_{\phi}P(\phi+\epsilon) & = \lim_{\Delta\rightarrow0} \frac{1}{\Delta} \Big( P(\phi+\epsilon+\Delta)-P(\phi+\epsilon) \Big)\nonumber \\
 & = \lim_{\Delta\rightarrow0} T_{\epsilon} \frac{1}{\Delta} \left( P(\phi+\Delta)-P(\phi) \right) T_{\epsilon}^{-1}\nonumber \\
 & = T_{\epsilon} \partial_{\phi}P(\phi) T_{\epsilon}^{-1}\,.
\end{align}

The Chern number associated with the gap is given by~\cite{Avron0Appendix}
\begin{align} \label{Eq:Chern}
\nu & =\frac{1}{2\pi i}\int_{0}^{2\pi}d\phi d\theta\, C(\theta,\phi)\,,
\end{align}
 where
\begin{align}
C(\phi,\theta) & =\textrm{Tr}\left(P\left[\frac{\partial P}{\partial\phi},\frac{\partial P}{\partial\theta}\right]\right)\,,\label{Eq:Chern-1}
\end{align}
is the Chern density. Notably,
\begin{align}
C(\phi+\epsilon) & = \textrm{Tr} \left(\left[P(\phi+\epsilon),\left[\partial_{\phi}P(\phi+\epsilon),\partial_{\theta}P(\phi+\epsilon)\right]\right]\right)\nonumber \\
 & =\textrm{Tr}\left(T_{\epsilon}\left[P(\phi),\left[\partial_{\phi}P(\phi),\partial_{\theta}P(\phi)\right]\right]T_{\epsilon}^{-1}\right)\nonumber \\
 & =C(\phi)\,.\label{Eq:Chern_insensitive}
\end{align}
 We can see that $C(\phi)$ is periodic with $1/L$ periodicity, which
means that
\begin{align}
\int_{0}^{2\pi}d\phi\, C(\phi) & =L\int_{0}^{2\pi/L}d\phi\, C(\phi)\nonumber \\
 & =2\pi C(\phi=0)+O(1/L)\,.
\end{align}
This implies that the Chern density, $C$, is independent of
$\phi$. Therefore, the integration over $\phi$ in Eq.~\eqref{Eq:Chern} is redundant,
and the Chern number is given by the Chern density at a given $\phi$. We can
therefore conclude that while formally the Chern number is associated with the
entire family $\{H(\phi)\}_{\phi=0}^{2\pi}$, we can actually associate the Chern
number with any member of the family $H(\phi)$.

The generalization of this proof to other QCs is straightforward. The proof relies
only on the periodicity of the Hamiltonian in $\phi$ and on the equivalence between
phase shifts and translations. Given a Hamiltonian of a $D$-dimensional QC with $d$
quasiperiodic terms, which can be associated with phase shifts, one can define a
periodic version of the Hamiltonian (similar to $H_{b}$). The same conversion from
phase shifts to translations still holds, as well as the insensitivity to the
periodicity and phase twists in the thermodynamic limit. Therefore any gap of such a
$D$-dimensional Hamiltonian is characterized by the topological classification of a
Hamiltonian of $D+d$ dimensions.

It is worth mentioning that the above argument does not imply that the phase $\phi$
has no physical meaning. Shifting $\phi$, or equivalently translating the lattice,
changes considerably the wavefunctions but does not alter the energies nor the Chern
density. This can be seen by the fact that the Chern density is defined as a
response to infinitesimal changes in $\phi$, but this response does not depend on
the value of $\phi$.

As a last note, it can be similarly shown that, in the thermodynamic limit the
energies and the Chern density does not depend on phase twist, $\theta$, as well.
The proof is the same as the one used for the IQHE~\cite{Niu_Thouless,ZoharKobiAppendix}.
Therefore, the periodic geometry serves only as a formal tool for defining the Chern
density but, in fact, it is geometry-free bulk property.


\section{Further theoretical discussion}
\label{Sec:FAQ}

This section addresses several theoretical issues that are mentioned in the main
text and deserve additional discussion for the benefit of the interested reader.

\subsection{Appearance of boundary states}

In the main text we state that any given QC, namely $H(\phi)$, is associated with a
Chern number and belongs to a nontrivial topological phase. On the other hand, in
order to witness the topological gap-traversing modes it is necessary to scan
$\phi$, i.e.~for a given system there is either one or no such boundary state on
each boundary. This seems to contradict the conventional notion that topological
phases, especially those with nontrivial Chern numbers, are always accompanied by
topologically protected gap-traversing boundary states. While this is often true,
there are examples for topological phases that do not necessarily have protected
boundary states.

Consider topological phases that stem from a non-local symmetry such as
inversion~\cite{Pollmann} or rotation~\cite{Fu_Crystal}. In such systems the
boundaries may not satisfy the symmetry of the bulk, and therefore may not host
sub-gap boundary states. QCs do not rely on symmetry, but on quasiperiodicity.
Strictly speaking, the boundary breaks the quasiperiodicity, hence may not have
boundary states. However, the mapping to 2D guarantees that by scanning $\phi$ these
states must appear.

Furthermore, in 1D topological phases the boundary states do not traverse the bulk
gap even in the thermodynamic limit. This is due to the fact that in 1D systems the
boundary does not scale with the system size. Therefore at the interface between
distinct topological phases the number of sub-gap states is of $O(1)$, e.g.~the
appearance of a Majorana fermion in 1D topological
superconductors~\cite{Kitaev_Majorana}. Nevertheless, since our system inherits its
topological behavior from a 2D system, the boundary states indeed traverse the bulk
gap, but only when $\phi$ is scanned.

\subsection{Quasicrystal vs.~crystal}

In our analysis we focused on quasiperiodic systems. One may ask what happens to the
topological behavior in a periodic system. The quasiperiodicity is characterized by
the irrational number, denoted above by $b$. This $b$ can always be approximated by
some rational number $p/q$, where $p$ and $q$ are integers. Such an approximation
will result in a periodic system. A general property of QCs is that this
approximation differs from the irrational case by $O(1/q)$ effects, as we will now
demonstrate.

For the irrational $b$ the spectrum is fractal~\cite{HofstadterAppendix}. Hence, it contains
gaps of all scales. Conversely, for the rational $p/q$ the spectrum is composed of
$q$ bands~\cite{Harper}. Therefore, taking the rational approximation will cause the
gaps with energy smaller than (band width)$/q$ to close.

Moreover, the independence of the energies on $\phi$, that is shown in the main
text, relies on the equivalence between shifts in $\phi$ and translations of the
lattice. This equivalence holds only for the irrational case. For $p/q$ there are
only $q$ values of $\phi$-shifts that are equivalent to translations. Therefore in
the rational case the spectrum varies with $\phi$ and is only periodic $2\pi/q$.

Recalling the proof in Section \ref{Sec:Chern}, it is evident that in the rational
case also the Chern density depends on $\phi$ with $2\pi/q$ periodicity. Therefore,
in order to evaluate the Chern number one has to perform the integration over
$\phi$. This means that the Chern number is associated only with the whole family of
1D systems, and that a single periodic system belongs to the trivial phase. This can
be also understood by the fact that deformation of such a family into a trivial
family will cause the bulk gap to close only for $q$ values of $\phi$, while for the
other values it remain open. Nevertheless, in case that the whole family of
Hamiltonians is nontrivial, boundary states will indeed traverse the bulk gap as a
function of $\phi$.

Finally, we note that for a finite system of length $L$, it will be physically
impossible to distinguish between an irrational $b$ and a rational approximation
$p/q$ for $q \gg L$.

\subsection{Robustness}

Topological phases are characterized by having their properties being robust to
perturbations. This is also the case for QCs. The topological classification of QCs
holds also in the presence of disorder, as long as the disorder preserves the
quasiperiodicity and does not close the bulk gap. Therefore (i) continuous
deformation between two topologically distinct systems results in a phase transition
also in disordered systems; (ii) The appearance of boundary states that traverse the
gap as a function of $\phi$ is also guaranteed. Since, any experimental setup is
inherently disordered to some level, this property was also demonstrated in our
experiments.

Considering fermionic systems, temperature should also be taken into account.
Since temperature only determines the occupation, the topological phase is stable to
temperatures that are smaller than the bulk gap. Similarly, when interactions are
introduced a quantized Chern number is still expected to hold as long as the bulk
gap remains open.

\subsection{Alternative gedanken experiment}

The association of a Chern number with a given QC, together with the above
discussions, can be demonstrated in the following gedanken experiment. Suppose we
have a sample of a 1D QC of finite but large length that is populated with electrons
up to some chemical potential. The sample should be put under an STM, which measures
its local density of states (LDOS). In general, we will find that the spectrum is
gapped over the bulk, whereas at each boundary a sub-gap boundary state may or may
not appear. If there are no boundary states, can we conclude that the QC is trivial?

The answer is no. We should cut the very last piece of the lattice at one of the
boundaries, and repeat the measurement. The gap at the bulk will, of course, remain
the same, but a sub-gap boundary state will probably appear. We repeat this process
several times. If the state is topological, then during repeated cuts the state will
disappear and reappear, but each time at a different energy. This happens since
cutting of the last piece can be thought of as a shift in $\phi$.

Integrating the LDOS at the boundary measured during this process, the gap will be
gradually filled. Notably, given the quasiperiodicity $b$, each cut is equivalent to
a shift of $2\pi b$. Therefore, by restoring the sub-gap spectrum as a function of
$\phi$ the Chern number can also be resolved from the measured data.

At this point, one should notice that in case of a periodic system with frequency
$p/q$, boundary states will also disappear and reappear, but now their energy will
have only $q$ different values. Hence, the integrated LDOS within the gap will be
composed of $q$ peaks, rather than being continuous.

The described experiment promotes several observations: First, It is performed on a
single member of the QC family with open boundaries. We can therefore see that it is
a bulk property and that the periodic geometry that was introduced in Section
\ref{Sec:Chern} is merely a formal tool. Second, if we take another member of the
family (i.e. a sample with the same $b$ but a different $\phi$) at the same chemical
potential and use this protocol, we would extract the same Chern number. This
demonstrates the independence of the Chern density on $\phi$. Third, the boundary
states that fill the gap in the integrated LDOS originate from topological edge
states in 2D. Last, it is evident that the physics is robust to disorder and
temperature that does not close the bulk energy gap.


\section{Generalization to other types of quasicrystals}
\label{Sec:Fibonacci}

In the main body of the paper we considered two examples of 1D quasicrystals that
exhibit topological properties: (i) the ``diagonal'' Aubry-Andr\'{e} model
[cf.~equation (1)] in which the quasiperiodicity is in the on-site potential, and
(ii) its ``off-diagonal'' version [cf.~equation (2)] that has its quasiperiodicity
in the hopping term. We then  argued that  tight-binding Hamiltonians of other QCs
and of higher dimensions also have such topological properties.

However, there is a class of QCs that is obtained by a ``cut and project''
method~\cite{QC_JanotAppendix,QC_Steinhardt}, for which one might think that our argument
does not hold. For these QCs the additional d.o.f.~are shifts of the ``cut'' in the
higher-dimensional space. Such shifts modify the Hamiltonian in a discontinuous
manner.  Hence, as our argument relied on the continuity of the on-site and hopping
terms as a function of the additional d.o.f.~, it seems that these QCs do not fit
our framework.

We now show that our generalization indeed includes such QCs. This is done by defining
a smooth version of the projection procedure. This procedure keeps the energy gaps
unchanged, and therefore has the same topological properties.
For example, in 1D, a QC can be obtained by projecting a square lattice
on the line $y=x/\tau$, where $\tau$ is irrational. The normalized
spacings between the lattice sites $n$ and $n+1$ is therefore
\begin{align}
d_n = 1 + \left( \tau-1 \right) \left( \left\lfloor
\left(n+2\right)\frac{\tau}{1+\tau} \right \rfloor - \left\lfloor
\left(n+1\right)\frac{\tau}{1+\tau}\right\rfloor \right)\,.
\end{align}
The case of $\tau=\left(1+\sqrt{5}\right)/2$ is the well studied Fibonacci
QC~\cite{QC_JanotAppendix,QC_Steinhardt}. A smoothed projection can be of the form
\begin{align}
&\bar{d}_{n}\left(\beta,\phi\right)  =1+\frac{\tau-1}{2}\times\\ &\biggl(1-\textrm{tanh}\beta\left[\sin\left(2\pi\left[\frac{2n+3}{2\tau}-\frac{1}{4}\right]+\phi\right)+\cos\frac{\pi}{\tau}\right]\biggr)\,.\nonumber
\end{align}
It can be easily shown that
$\lim_{\beta\rightarrow\infty}\bar{d}_{n}\left(\beta,\phi=0\right)=d_{n}$. Note that
$\phi$ is embedded into $\bar{d}_{n}$, in a way that is consistent with our general
argument.

The spacing between the lattice sites enters the Hamiltonian through the hopping
terms. Since a Hamiltonian with $\bar{d}_{n}$ can be continuously deformed into one
with $d_{n}$ without closing the energy gaps, they belong to the same topological
class. Therefore, given a QC which is obtained by a ``cut and project'' procedure,
one should find a smoothed version of the projection and accordingly obtain a
smoothed Hamiltonian. Now, the topological properties of the ``cut'' Hamiltonian can
be deduced from the smoothed one. Note, that for ``cut and project''
quasiperiodicity which is incorporated in the on-site terms, a similar smoothing can
be performed. Thus, we have shown that the ``cut and project'' QCs also fit into our
general argument.


\begin{thebibliography}{100}
\bibitem{pumping}
This is analogous to Laughlin's pumping in the IQHE (see Ref.~\cite{Laughlin}).
However, here a single mode is being pumped, while in the IQHE all the electrons of
the occupied band contribute to the current and induce the charge pumping.
\end{thebibliography}
\end{document}